\input{aipcheck}

\documentclass[
    ,final            
  ]
  {aipproc}

\layoutstyle{6x9}

\begin{document}

\title{Autonomous Dynamics in Neural networks: The
dHAN Concept and Associative Thought Processes}


\author{Claudius Gros}{
  address={Institute for Theoretical Physics
J.W. Goethe University Frankfurt, Germany.}
}



\begin{abstract}
The neural activity of the human brain is
dominated by self-sustained activities.
External sensory stimuli influence
this autonomous activity but they do not
drive the brain directly. Most standard artificial
neural network models are however input driven
and do not show spontaneous activities. 

It constitutes
a challenge to develop organizational principles for
controlled, self-sustained activity in artificial
neural networks. Here we propose and examine the
dHAN concept for autonomous associative thought processes
in dense and homogeneous associative networks.
An associative thought-process is characterized, within
this approach, by a time-series of transient attractors. Each 
transient state corresponds to a stored information,
a memory. The subsequent transient states are
characterized by large associative overlaps,
which are identical to acquired patterns. Memory states,
the acquired patterns, have such a dual functionality.

In this approach the self-sustained neural activity
has a central functional role. The network acquires a
discrimination capability, as external stimuli need
to compete with the autonomous activity. Noise in
the input is readily filtered-out.

Hebbian learning of external patterns occurs coinstantaneous  
with the ongoing associative thought
process. The autonomous dynamics needs a long-term
working-point optimization which acquires within the
dHAN concept a dual functionality: It stabilizes the
time development of the associative thought process 
and limits runaway synaptic growth, which generically
occurs otherwise in neural networks with self-induced activities and
Hebbian-type learning rules.
\end{abstract}

\maketitle


\section{Cognitive system theory}

The present approach is situated within the general framework
of cognitive system theory. Let us start with a general
definition of a cognitive system.
\begin{quotation}
\noindent
\underline{\sl Cognitive systems}\\
A cognitive system is a continuously active
complex adaptive system autonomously
exploring and reacting to the environment
with the capability to `survive'.
\end{quotation}
A cognitive system is an abstract dynamical system.
It might be either biological or cybernetical. Our
brain, to give an example, is the physical support
of the human cognitive system. A cognitive system `dies'
whenever its physical support looses functionality.

The condition for `survival' can be phrased in a mathematical
precise way. The physical support of a cognitive system
remains functional only when a set of key parameters remain
within a given range. Examples for such `survival parameters' 
are the blood-sugar level for a biological cognitive system 
or the battery status for an autonomous robot.
The cognitive system receives information about the status
of these survival parameters via appropriate sensors. It
survives if its activity keeps its physical support functional
via appropriate motor outputs.

It is desirable to develop artificial cognitive systems
which could operate in a wide range of possible real-world
or simulated environments, {\it viz.} cognitive systems
with universal capabilities which are not tailored for
specific task solutions. There are then several important
points to be taken into account.
\begin{enumerate}
\item\underline{Autonomous dynamics}\\
The human brain dynamics is dominantly self-sustained.
It is influenced but not driven by the sensory input 
\cite{Fiser04}. It is therefore necessary to propose and
to study possible organizational principles for neural
networks with self-sustained dynamical activities.

\item\underline{Homeostatic principles}\\
The brain adapts itself autonomously to a wide
range of growth conditions and injuries. It is
therefore of interest to study neural-network layouts
which regulate most parameters, as synaptic strength
and learning rates, homeostatically.

\item\underline{Unsupervised learning}\\
Most learning by an autonomous cognitive system should
be unsupervised - the system selects when and what to
learn. Learning rules should be local, such that the 
system is scalable and remains functional under 
structural modifications.

\item\underline{Online learning}\\
There should be no distinct phases for learning
and performance. Learning should be `on-the-fly'.

\item\underline{Universality}\\
The layout principles for the cognitive system should be based,
as far as possible, on universal principles. A priori knowledge 
about the environment can be added in a second step, if necessary,
in order to boost the performance for specific tasks.
\end{enumerate}
%

\begin{figure}
  \includegraphics[width=.6\textwidth]{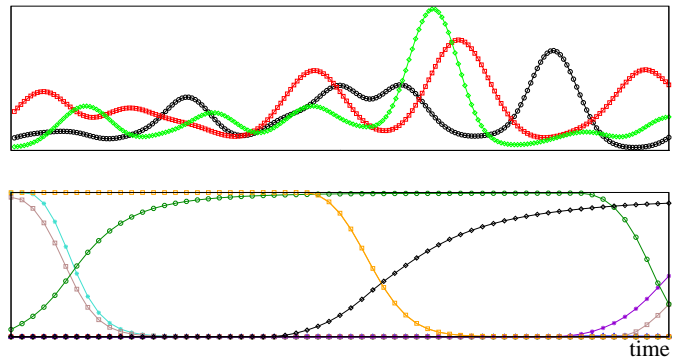}
  \caption{Illustration of fluctuating (top) and transient-state
           dynamics (bottom).}
\label{fig_trans_states}
\end{figure}

\section{Basic cognitive system theory principles}

In addition to the rather general principles 
stated above one can formulate, guided by the
results of neurobiological studies, several
important guidelines.
\begin{itemize}
\item\underline{Competitive brain dynamics}\\
Studies of the neural correlate for conscious cognitive states 
suggest the formation of `winning coalitions', also called 
`critical reentrant events' \cite{Edelman03}, of 
competitively active neural ensembles. This competitive
brain dynamics takes place in what is called a
`global workspace' \cite{Dehaene03}, made-up of
essential nodes \cite{Crick03}.

\item\underline{Transient-state dynamics}\\
Competitive dynamics naturally results in
transient state dynamics, see Fig.\ \ref{fig_trans_states},
as the winning coalition of neural ensembles suppresses
the activities of competing centers. A time series of semi-stable 
winning coalitions of computational subunits then results.

\item\underline{Autonomous brain dynamics}\\
The spontaneously generated neural activity patterns
generated in the cortex are not void of contextual
information. It has been observed, that they resemble
(in the visual cortex) memories of previous visual stimuli
\cite{Ringach03,Kenet03}, forming transient states
\cite{Abeles95}.

\item\underline{Associative thought processes}\\
Humans dispose over a huge commonsense database,
mostly organized associatively \cite{Nelson98,Liu04}.
It is therefore reasonable to assume, that the
autonomous dynamics reflects this fact. A possible
paradigm for the self-sustained dynamics, which we will 
follow here, is than that of associative thought processes.
Subsequent transient states then correspond to memories
connected associatively.

\item\underline{Sparse coding}\\
Neural networks with sparse coding\footnote{Sparse
coding is present in a neural network when, on the average,
only a small fraction of all neurons is active simultaneously.}
have a storage capacity orders of magnitude larger than networks
with an average activity level of 50\% \cite{Arbib02}.
Clique\footnote{In network theory one denotes by `clique'
a fully interconnected subcluster.} encoding, an
instance of a `winners-take-all' encoding, combines
competitive dynamics with the large storage capacity of
sparse coding. For clique encoding, which we will consider
here, the winning coalitions are constituted by mutually
supporting neural activity centers. The notion of
clique encoding also draws from studies in cognitive science
indicating the importance of the 
`chunking mechanism'\footnote{Chunking denotes the notion of 
grouping together elementary units of information for memory
formation} for human learning \cite{Gobet01,Boucher03}.
\end{itemize}
The study of neural networks which incorporate above
principles is hence a necessary step towards the
eventual development of an artificial cognitive system.
Here we present and study a generalized neural network 
which implements these requirements necessary for any 
component of an autonomous cognitive system. The network 
we examine is suitable to store patterns occurring
in the sensory data input stream via unsupervised
learning.

\begin{figure}[t]
\centerline{
\includegraphics[height=0.30\textwidth]{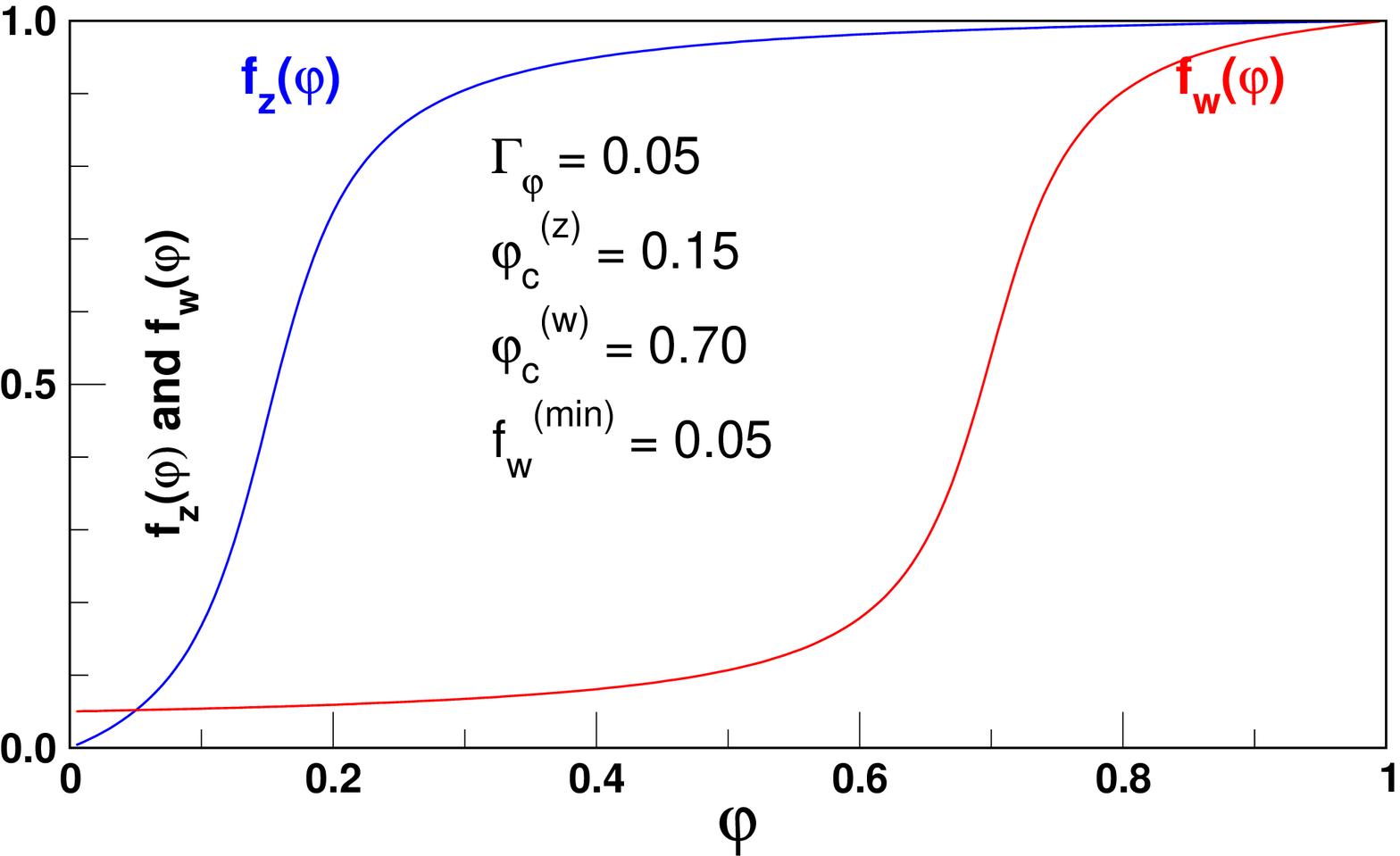}
\hspace{6ex}
\includegraphics[height=0.30\textwidth]{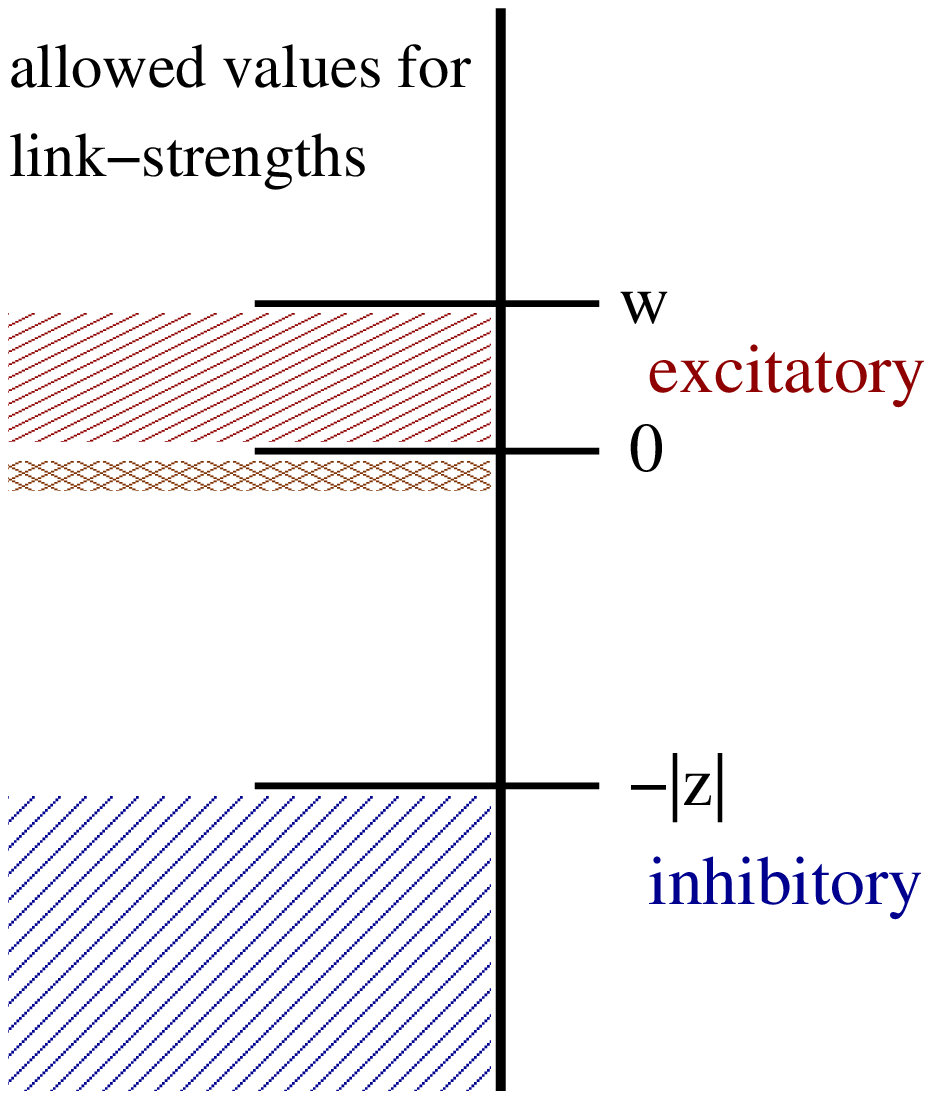}
           }
\caption{Left: Illustration of the reservoir functions
$f_{z/w}(\varphi)$, see Eq.\ \ref{cogSys_ri},
of sigmoidal form with respective turning points
$\varphi_c^{(f/z)}$, a width $\Gamma_\varphi$ and a minimal
value $f_z^{(min)}=0$. \newline
Right: Distribution of the synapsing strength leading to
clique encoding. Weak inhibitory synapsing 
strengths do not occur.
}
\label{cogSys_fig_gaps}
\end{figure}

\section{Associative thought processes}

\begin{figure}[t]
\centerline{
\includegraphics*[height=0.40\textwidth]{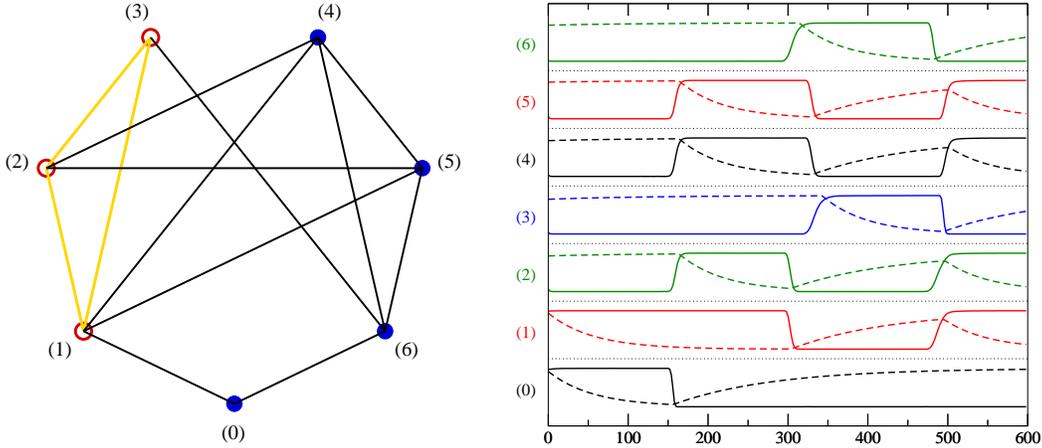}\hspace{3ex}
\includegraphics*[height=0.40\textwidth]{AI7.eps}
           }
\caption{Left: A 7-site network,
shown are links with $w_{i,j}>0$, containing six cliques,
(0,1), (0,6), (3,6), (1,2,3) (which is highlighted), 
(4,5,6) and (1,2,4,5). \newline
Right: 
The activities $x_i(t)$ (solid lines)
and the respective reservoirs $\varphi_i(t)$ (dashed lines)
for the transient-state dynamics $(0,1)\rightarrow(1,2,4,5)
\rightarrow (3,6) \rightarrow(1,2,4,5)$.
\label{fig_7sites}
        }
\end{figure}

We now present an implementation, in terms
of a set of appropriate coupled differential equations,
of the notion of associative thought processes as a
time series of transient attractors. We start by a network
with a large number of stable attractors (the cliques)
and then turn these attractors into transient attractors
by coupling to a second variable with a long
time scale. To be concrete, we denote with
$x_i\in[0,1]$ the activities of the local computational
units constituting the network and with
$ \varphi_i\ \in\ [0,1] $
a second variable which we call {\em `reservoir'}.
The differential equations \cite{Gros05}
\begin{eqnarray} \label{cogSys_xdot}
\dot x_i &=& (1-x_i)\,\Theta(r_i)\,r_i \,+\, x_i\,\Theta(-r_i)\,r_i 
\\ \label{cogSys_ri} 
    r_i &=& 
    \sum_{j=1}^N \Big[
    f_w(\varphi_i) \Theta(w_{ij}) w_{i,j}
    + z_{i,j}f_z(\varphi_j)
        \Big] x_j\ \ \
\\ \label{cogSys_phidot}
\dot\varphi_i & =&  
\Gamma_\varphi^+\, (1-\,\varphi_i)(1-x_i/x_c)
\Theta(x_c-x_i)
\,-\, \Gamma_\varphi^-\,\varphi_i\,\Theta(x_i-x_c)
\\ \label{cogSys_z_t}
z_{ij} & =& -|z|\,\Theta(-w_{ij})
\end{eqnarray}
generate associative thought processes. We now
discuss some properties of (\ref{cogSys_xdot}-\ref{cogSys_z_t}).
\begin{itemize}

\item \underline{Normalization}\\
Eqs.\ (\ref{cogSys_xdot}-\ref{cogSys_phidot}) respect the 
normalization $x_i,\varphi_i\in[0,1]$, due to
the prefactors $x_i$,$(1-x_i)$, $\varphi_i$
and $(1-\varphi_i)$
in  Eqs.~(\ref{cogSys_xdot}) and (\ref{cogSys_phidot}),
for the respective growth and depletion processes.
$\Theta(r)$ is the Heaviside-step function: 
$\Theta(r<0)=0$  and $\Theta(r>0)=1$.

\item\underline{Synapsing strength}\\
The synapsing strength is split into an
excitatory contribution $\propto w_{i,j}$ 
and an inhibitory contribution $\propto z_{i,j}$,
with $w_{i,j}$ being the primary variable:
The inhibition $z_{i,j}$ is present only
when the link is not excitatory (\ref{cogSys_z_t}).
With $z\equiv-1$ one sets the inverse unit of time.

\item\underline{Winners-take-all network}\\
Eqs.~(\ref{cogSys_xdot}) and (\ref{cogSys_ri}) describe,
in the absence of a coupling to the reservoir via 
$f_{z/w}(\varphi)$, a competitive winners-take-all 
neural network with clique encoding.
The system relaxes towards
the next attractor made up of a clique 
of $Z$ sites $(p_1,\dots,p_Z)$ connected via 
excitatory $w_{p_i,p_j}>0$ ($i,j=1,..,Z$). 

\item\underline{Reservoir functions}\\
The reservoir functions $f_{z/w}(\varphi)\in[0,1]$ 
govern the interaction in between the activity levels $x_i$ 
and the reservoir levels $\varphi_i$. 
They may be chosen as washed out step functions
of sigmoidal form 
with a suitable width $\Gamma_{\varphi}$ and
inflection points $\varphi_c^{(w/z)}$, see
Fig.\ \ref{cogSys_fig_gaps}.

\item\underline{Reservoir dynamics}\\
The reservoir levels of the winning clique 
depletes slowly, see Eq.~(\ref{cogSys_phidot})
and Fig.\ \ref{fig_7sites}, and recovers only
once the activity level $x_i$ of a given site has
dropped below $x_c$.
The factor $(1-x_i/x_c)$ occurring in
the reservoir growth process, see the r.h.s.\ of
(\ref{cogSys_phidot}), serves for a stabilization of the
transition between subsequent memory 
states \cite{Gros05}

\item\underline{Separation of time scales}\\
A separation of time scales is obtained when the
$\Gamma_\varphi^\pm$ are much smaller than the
average strength of an excitatory link, $\bar w$, 
leading to transient-state dynamics. 
Once the reservoir of a winning clique
is depleted, it looses, via $f_z(\varphi)$, its ability to 
suppress other sites and the mutual intra-clique
excitation is suppressed via $f_w(\varphi)$. 
\end{itemize}

In Fig.\ \ref{fig_7sites} the
transient-state dynamics resulting from
Eqs.\ (\ref{cogSys_xdot}-\ref{cogSys_z_t}),
in the absence of any sensory signal, is illustrated.
When the growth/depletion rates $\Gamma_\varphi^\pm\to0$
are very small, the individual cliques turn into stable
attractors. 
The possibility to regulate the `speed' of the
associative thought process arbitrarily by setting
the $\Gamma_\varphi^\pm$ is important
for applications. For a working cognitive system
it is enough if the transient states are just stable
for a certain minimal period, anything longer just
would be a `waste of time'.

\paragraph{Cycles}
The system in Fig.\ \ref{fig_7sites} is
very small and the associative thought process
soon settles into a cycle, since there are
no incoming sensory signals in the simulation 
of Fig.\ \ref{fig_7sites}.
For networks containing a somewhat larger number of sites, the
number of attractors can be however very large and such
the resulting cycle length. We performed simulations
for a 100-site network, to give an example, containing
713 clique-encoded memories. We found no cyclic behavior
even for thought processes with up to 4400 transient states.
For a working cognitive system prolonged periods 
without sensory signals will be anyhow rare events 
and it will be unlikely that the system will settle 
into a stable cycle of memories.

\paragraph{Dual functionalities for memories}
The network discussed here is a dense and
homogeneous associative network (dHAN).
It is homogeneous since memories have dual functionalities:
\begin{itemize}
\item Memories are the transient states of the associative thought
      process.
\item Memories define the associative overlaps, between two
      subsequent transient states.
\end{itemize}

\paragraph{Recognition}
Any sensory stimulus arriving to the dHAN needs to compete with
the ongoing intrinsic dynamics to make an impact. If the
sensory signal is not strong enough, it cannot deviate
the autonomous thought process. This feature results
in an intrinsic recognition property of the dHAN: A background
of noise will not influence the transient state dynamics.

\section{Autonomous online learning}

An external stimulus, $\{b_i^{(ext)}(t)\}$,
influences the activities $x_i(t)$ of the respective neural
centers. This corresponds to a change of the respective
growth rates $r_i$,
\begin{equation}
r_i \ \to \ r_i \,+\,f_w(\varphi_i)\,b_i^{(ext)}(t)~,
\label{cogSys_stim}
\end{equation}
compare Eq.\ (\ref{cogSys_ri}), where $f_w(\varphi_i)$
is an appropriate coupling function, depending on the
local reservoir level $\varphi_i$. The task is then
to formulate principles which let the dHAN learn and
store on-thy-fly patterns found in the stimuli
$b_i(t)$.

\subsection{Short- and long-term synaptic plasticities}

There are two fundamental considerations for the choice
of synaptic plasticities adequate for the dHAN.
\begin{itemize}
\item Learning is a very slow process without a short-term memory.
      Training patterns need to be presented to the network over and
      over again until substantial synaptic changes are induced \cite{Arbib02}.
      A short-term memory can speed-up the learning process substantially
      as it stabilizes external patterns and hence
      gives the system time to consolidate long-term
      synaptic plasticity. 

\item Systems using sparse coding are based on a strong
      inhibitory background, the average inhibitory link-strength
      $|z|$ is substantially larger than the average
      excitatory link strength $\bar w$,
$$
  |z|\ \gg\  \bar w~.
$$
      It is then clear that gradual learning affects dominantly 
      the excitatory links: Small changes of large 
      parameters do not lead to new transient
      attractors, nor do they influence the cognitive dynamics
      substantially.
\end{itemize}
We then have
\begin{equation}
w_{ij}\ =\  w_{ij}(t)\ =\  w_{ij}^{S}(t)\,+\, w_{ij}^{L}(t)~,
\label{cogSys_w_S_L}
\end{equation}
where $w_{ij}^{S/L}$ correspond to the short/long-term 
synaptic plasticities. 

\paragraph{Negative baseline}
Eq.\ (\ref{cogSys_z_t}), $z_{ij} = -|z|\,\Theta(-w_{ij})$, 
states that the inhibitory link-strength is either
zero or $-|z|$, but is not changed directly during learning,
in accordance to (\ref{cogSys_w_S_L}).
When a $w_{i,j}$ is slightly negative, as default
(compare Fig.\ \ref{cogSys_fig_gaps}),
the corresponding total link strength is inhibitory.
When $w_{i,j}$ acquires, during learning, a positive value,
the corresponding total link strength becomes excitatory.
\begin{figure}[t]
\centerline{
\includegraphics[width=0.40\textwidth]{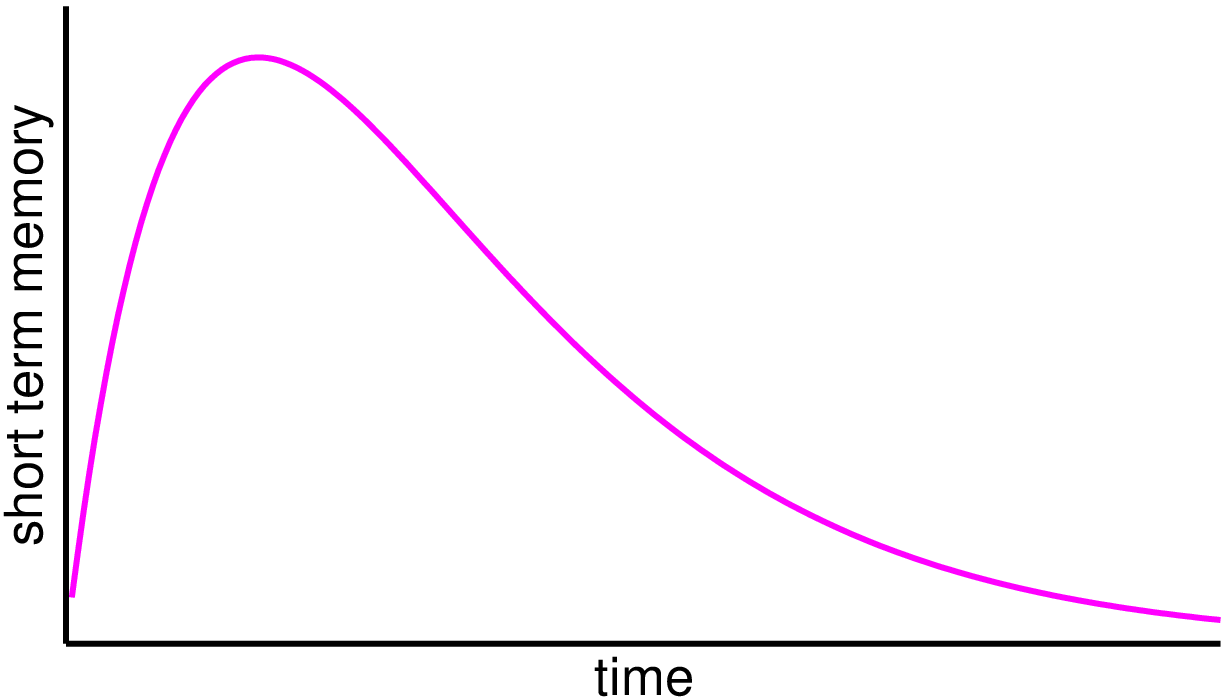}
\hspace{1ex}
\includegraphics*[width=0.55\textwidth]{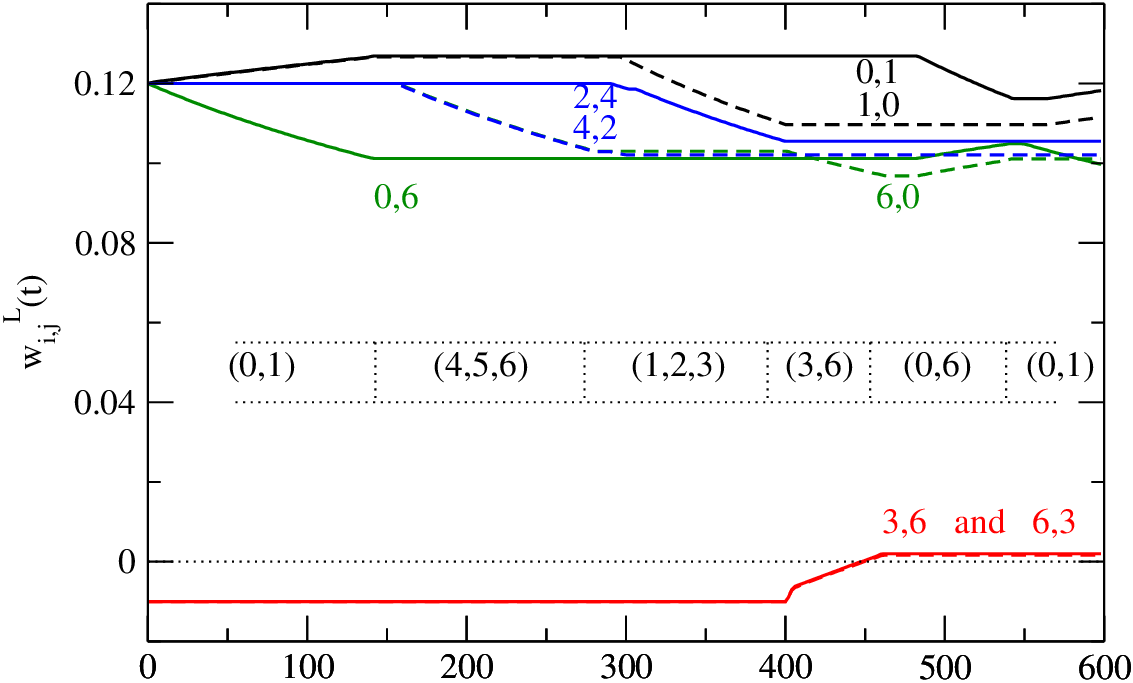}
           }
\caption{Left: Typical activation pattern of the short-term
plasticities of an excitatory link (short-term memory). \newline
Right: The time evolution of the long-term memory, 
for some selected links $w_{i,j}^L$ and the network illustrated
in Fig.\ \ref{fig_7sites}, without the link (3,6).
The transient states are
$(0,1)\rightarrow(4,5,6)
\rightarrow(1,2,3)
\rightarrow(3,6)
\rightarrow(0,6)
\rightarrow(0,1)
$.
An external stimulus at
sites (3) and (6) acts for $t\in[400,410]$ with
strength $b^{(stim)}= 3.6$. The stimulus pattern
(3,6) has been learned by the system, as the
$w_{3,6}$ and $w_{6,3}$ turned positive during the
learning-interval $\approx [400,460]$. The learning
interval is substantially longer than the bare
stimulus length due to the activation of the short-term
memory.
 }
\label{cogSys_fig_memory}
\end{figure}

\subsection{Short-term memory dynamics}

It is reasonable to have a maximal possible value
$W_S^{(max)}$ for the short-term synaptic
plasticities. The appropriate Hebbian-type 
autonomous learning rule is then
\begin{eqnarray}
\label{cogSys_w_S_dot}
\dot w_{ij}^S(t) & =&
\Gamma_{S}^+
\left(W_S^{(max)}-w_{ij}^S\right)
f_z(\varphi_i) f_z(\varphi_j) \,
\Theta(x_i-x_c) \Theta(x_j-x_c) \\
&-& \Gamma_{S}^-\,w_{ij}^S~.
\nonumber
\end{eqnarray}
It increases rapidly when both the pre- and 
the post-synaptic centers are active, it decays
to zero otherwise, see Fig.\ \ref{cogSys_fig_memory}.
The coupling functions $f_z(\varphi)$ preempt 
prolonged self-activation of the short-term
memory. When the pre- and the post-synaptic centers 
are active long enough to deplete their respective
reservoir levels, the short-term memory is shut-off
via $f_z(\varphi)$, compare  
Fig.\ \ref{cogSys_fig_gaps}.

\begin{figure}[t]
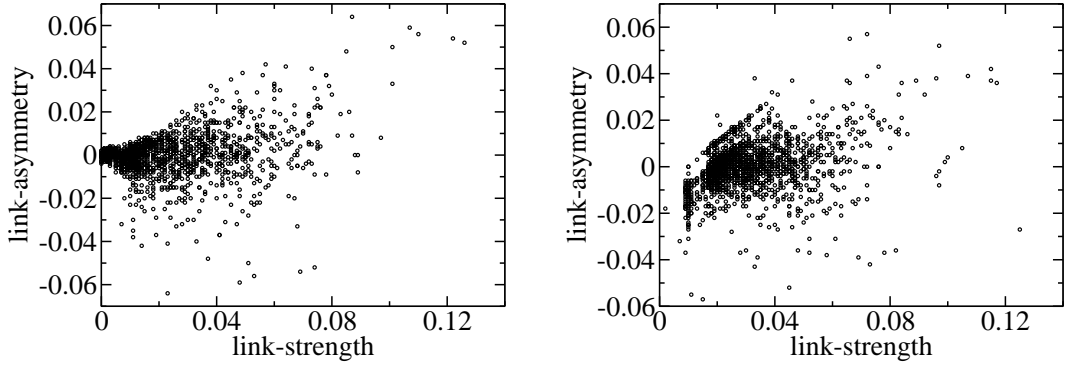

\centerline{
\includegraphics*[width=0.45\textwidth]{100WW_learn.eps}
\hspace{3ex}
\includegraphics*[width=0.45\textwidth]{100WW_start.eps}
           }
\caption{The link-asymmetry $w_{ij}^L-w_{ji}^L$ 
for the positive $w_{ij}^L$ 
for a 100-site network with 713 cliques
at time $t=500\, 000$, corresponding to circa
4500 transient states.
\newline
Left: After learning from
scratch, learning was finished at $t\approx50\, 000$. \newline
Right: Starting with $w_{i,j}\to0.12$ for
all links belonging to one or more cliques.
\label{cogSys_fig_100_links}
        }
\end{figure}

\subsection{Long-term memory dynamics}

Dynamical systems retain normally their functionalities
only when they keep their dynamical properties
in certain regimes. They need to
regulate their own working point. This is
a long-term affair, it involves time-averaged
quantities, and it is therefore a job for the
long-term synaptic plasticities,
$w_{ij}^{L}$. 

\paragraph{Effective incoming synaptic strength}
The average magnitude of the 
growth rates $r_i$, see Eq.\ (\ref{cogSys_ri}),
determine the time scales of the autonomous dynamics
and such the working point.  The $r_i(t)$ 
are however quite strongly time dependent.
The effective incoming synaptic signal
\[
\tilde r_i \,=\, \sum_{j}\Big[ w_{i,j}x_j
  \,+\, z_{i,j}x_jf_z(\varphi_j)\Big]~, 
\]
which is independent of the post-synaptic
reservoir, $\varphi_i$, is a more convenient
control parameter. The working point of the cognitive
system is optimal when the effective
incoming signal is, on the average, of comparable
magnitude $t^{(opt)}$ for all sites,
\begin{equation}
\tilde r_i\ \to\ r^{(opt)}~.
\label{eq_r_to_r_opt}
\end{equation}
Eq.\ (\ref{eq_r_to_r_opt}) is an implementation of the
principle of homeostatic self-regulation.

The long-term memory has two tasks: To encode
the stimulus patterns and to keep the working
point of the dynamical system in its desired
range. Both tasks can be achieved by a single local
learning rule,
\begin{eqnarray}
\label{cogSys_w_L_dot}
\dot w_{ij}^L(t) & =& 
\Gamma_{L}^{(opt)} 
\Delta \tilde r_i \Big[\,
\left(w_{ij}^L-W_L^{(min)}\right) \Theta(-\Delta \tilde r_i) +
\Theta(\Delta \tilde r_i)
                \,\Big] \\ &&\cdot
\,\Theta(x_i-x_c)\,\Theta(x_j-x_c),
\qquad\qquad\qquad\qquad
\Delta \tilde r_i\,=\,r^{(opt)}-\tilde r_i~.
\nonumber
\end{eqnarray}
Some comments:
\begin{itemize}
\item\underline{Hebbian learning}\newline
      The learning rule is local and of Hebbian type \cite{Arbib02}. 
      Learning occurs only when the pre- and the post-synaptic
      neuron are active. Weak forgetting, i.e.\ the decay
      of seldom used links is not present in 
      (\ref{cogSys_w_L_dot}), but could be added to it.
\item\underline{Synaptic competition}\newline
       When the incoming signal is weak/strong, relative to
      the optimal value $r^{(opt)}$, the active links
      are reinforced/weakened, with $W_L^{(min)}$ being the
      minimal value for the $w_{ij}$. The baseline $W_L^{(min)}$
      is slightly negative, compare
      Figs.\ \ref{cogSys_fig_gaps} and \ref{cogSys_fig_memory}.
    
      The Hebbian-type learning then takes place in the form of 
      a competition between incoming synapses - frequently 
      active incoming links will gain strength, on the average, 
      on the expense of rarely used links. 

\item\underline{Fast learning of new patterns}\newline
      In Fig.\ \ref{cogSys_fig_memory} the time evolution
      of some selected $w_{ij}$ from a simulation is
      presented. A simple input-pattern is learned by the
      network. In this simulation the learning parameter
      $\Gamma_{L}^{(opt)}$ has been set to a quite large value
      such that the learning occures in one step (fast learning).

\item\underline{Suppression of runaway synaptic growth}\newline
      The link-dynamics (\ref{cogSys_w_L_dot}) suppresses
      synaptic runaway-growth, a general problem common to
      adaptive, continuously active neural networks. It has
      been shown that similar rules for discrete neural networks
      optimize the overall storage capacity \cite{Chechik01}.

\item\underline{Long-term dynamical stability}\newline
      In Fig.\ \ref{cogSys_fig_100_links} the results for the
      long-term link matrices are presented for a 100-site
      network with 713 stored memories and for two simulations.
\begin{itemize}
\item In the first simulation all excitatory links were 
      set by hand right at the start to 0.12 \cite{Gros05}. 
      The working-point
      optimization inherent in Eq.\ (\ref{cogSys_w_L_dot})
      then leads to a differentiation for the link-strengths
      during self-generated associative thought process,
      generating a total of about 4500 transient states.
\item In the second simulation all excitatory links were 
      learned on-the-fly, via Eqs.\ (\ref{cogSys_w_S_dot}) and
      (\ref{cogSys_w_L_dot}), from patterns presented
      to the network during $t\in[0,50\,000]$. Afterwards
      the dynamics was 100\% self-generated.
\end{itemize}
      The resulting final link distributions are similar. 
      This result indicates that self-sustained associative
      thought processes lead to stable long-term link distribution
      and such to stable cognitive dynamics. The system is
      self-adapting.
\end{itemize}

\section{Conclusions}

We have pointed out the importance of studying neural
networks layouts compatible with the requirements
for autonomously operating cognitive systems. We have
formulated a set of basic requirements and discussed
an implementation for a network capable to 
learn and store autonomously environmental data
as they occur in the sensory stimuli. 

We have pointed out (i) that fast online-learning is possible
when a short term memory complements the usual
long-term synaptic plasticities needed for 
pattern storage, (ii) that the working point
of the self-sustained dynamics can be regulated
homeostatically during the learning process
and (iii) that clique-encoding allows at the
same time for the generation of associative
thought processes and for a very high
storage capacity.







\IfFileExists{\jobname.bbl}{}
 {\typeout{}
  \typeout{******************************************}
  \typeout{** Please run "bibtex \jobname" to optain}
  \typeout{** the bibliography and then re-run LaTeX}
  \typeout{** twice to fix the references!}
  \typeout{******************************************}
  \typeout{}
 }



\end{document}